**Absence of superconductivity in LK-99 at ambient conditions**

Kapil Kumar[1,2], N. K. Karn[1,2], Yogesh Kumar[1,2] and V.P.S. Awana[1,2,*]

[1]*CSIR-National Physical Laboratory, Dr. K. S. Krishnan Marg, New Delhi-110012, India.*

[2]*Academy of Scientific and Innovative Research (AcSIR), Ghaziabad 201002, India.*

## Abstract:

The report of synthesis of modified Lead Apatite (LK-99), with evidence of superconductivity at more than boiling water temperature, has steered the whole scientific community. There have been several failures to reproduce superconductivity in LK-99, albeit partial successes. Here, we have continued our efforts to synthesize phase pure LK-99, with improved precursors. The process has been followed as suggested by Sukbae Lee *et. al.,*[1,2]. The phase purity of each precursor is evidenced by Powder X-ray diffraction (PXRD) and well fitted by Rietveld refinement. The PXRD confirms the synthesis of phase pure polycrystalline LK-99 with Lead Apatite structure. The freshly synthesized sample does not show any signature of superconductivity viz. levitation on a magnet or vice versa. The magnetization measurements on SQUID magnetometer show that LK-99 is diamagnetic at 280K, and there is no sign of superconductivity in LK-99 at room temperature. The sample is highly resistive as well. Moreover, we have also performed first principle calculations to investigate the electronic band structure of the LK-99 near Fermi level. Our study verifies that the Cu doped Lead Apatite (LK-99) has bands crossing at Fermi level, indicating generation of strong correlation in the system. Our results do not approve appearance of superconductivity in LK-99, i.e., $Pb_9CuP_6O_{25}$.

## *Corresponding Author

Dr. V. P. S. Awana: E-mail: awana@nplindia.org
Ph. +91-11-45609357, Fax-+91-11-45609310
Homepage: awanavps.webs.com

## Introduction:

The scientific community around the world is trying to reproduce the astonishing results of LK-99 (superconductivity near 400K) shown by the Korean group[1-3]. There have been two experimental reports[4,5] showing negative primary evidence, including ours[5]. A partial success in attaining superconductivity in the Cu doped Lead Apatite is shown by Q. Hou *et. al.,*[6]. But they showed superconductivity below 110K, which is way below the claimed superconductor transition temperature ($T_c$) of 400K. Another recent report shows magnetic levitation of LK-99 at room temperature with diamagnetic transition around 325K[7]. A close look in all attempts suggest that the phase purity of each precursor is very crucial for the synthesis of LK-99, and this must be ensured before expecting superconductivity in the sample. Two mechanisms for superconductivity in LK-99 have been proposed by the original creators of LK-99, one is Superconducting Quantum Well and second is Strong Correlation due to enhanced coulomb interaction by doping of Copper[1,2]. A quite good BR-BCS theory has been developed to predict transition temperature $T_c$ of high temperature superconductors[8] via strong electron correlation mechanism. Recently, Baskaran[9,10] has proposed an alternate mechanism, which suggests that there is a transition from broad band Mott insulator to superconducting region by introducing doping of Cu atom in Lead Apatite structure[9]. These two mechanisms[8,9] have opened up new unbound number of possible candidates of Hot superconductor to try out experimentally. Other computational studies[11-14] show that the parent Lead Apatite is a band gap insulator but the doping of Cu atom shows two isolated flat bands near Fermi level. On the other hand, the two-band Hubbard model on a triangular lattice developed by Hanbit Ho *et. al.*[15] suggests s-wave pairing. They show that LK-99 can be a possible low temperature superconductor, when the obtained DFT parameters for the two-band model are clubbed with Hubbard model[15]. Interestingly, another first principle based study[16] envisages that Cu-doping in LK-99 leads to ultra-flat bands crossing the Fermi level due to Cu(3d) and O(2p) orbitals hybridisation and possible superconductivity. Based on symmetry arguments, this is also envisaged[17] that Cu-doped LK-99 may lead to possible ferroic properties. Interestingly, all the results cited above are on cond-mat arxiv, and yet not been through the SCI journals rigorous peer review. This is because the room temperature superconductivity is evolving fast since after its breaking news[1,2].

This is our second attempt to find superconductivity in LK-99 as claimed by Sukbae Lee *et. al.,*[1,2]. This time, we have used high purity precursors. The same is evidenced by the synthesis of $Cu_3P$, where unreacted copper peaks are not found. In our previous report[5] as well as in partial success report[6], the PXRD of $Cu_3P$ showed the unreacted peaks of Cu. But this time there is no extra peak observed in PXRD of $Cu_3P$ following the same method. The phase purity of the synthesised LK-99 sample is confirmed by XRD measurement. In a most recent report[18], though the unreacted phase are visibly smaller than all earlier reports[1-3,4-6,] yet seems more than as of our present sample. To date our sample seems to be the purest one as far as its phase purity, to be discussed in results section. Our results support synthesis protocols of[1,2] for phase pure LK-99. The primary test for diamagnetism is magnetic levitation of the sample, which is not observed in our as synthesized sample. The magnetization measurement unequivocally demonstrates no superconductivity being present in the modified Lead Apatite at room temperature. The sample at room temperature is highly resistive as well. Our first principal studies support the mechanisms proposed by H. T. Kim[9], i.e., doping of Cu atoms at Pb site enhances the electronic correlations in LK-99.

## Experimental and Computational Details:

For the synthesis of target sample of LK-99, same protocol has been followed as in ref. [1,2,5]. Brief procedure is as follows: To synthesise first precursor $Cu_3P$, a freshly bought brown chunk of Phosphorous (P) from Sigma Aldrich is ground with Copper (Cu) powder in Argon filled M-Braun Glove box. The exposure to open atmosphere is strictly avoided. The ground powder is pelletized and vacuum sealed at $10^{-5}$ torr and heated for 48 hours at 550°C, as shown in Fig. 1(a). The resultant product encapsulated in quartz ampule after heat treatment is shown in Fig. 1(b). The Cu powder from Merck is pre-checked by PXRD to ensure the absence of CuO. Further, $PbSO_4$ and PbO being the essential elements for the production of other precursor Lanarkite $Pb_2SO_5$ are reacted. The chemical reaction that produces $PbSO_4$ is as follows; $Pb(NO_3)_2 + H_2SO_4 = PbSO_4 + 2HNO_3$.

The resultant, white colour powder is dried in oven, and its phase purity is validated using PXRD[5]. To synthesise Lanarkite, the freshly prepared $PbSO_4$ is mixed with high purity PbO (Sigma Aldrich) and the mixture is put in Alumina crucible. The Alumina crucible is subject to heat treatment as shown in Fig.1(c), in which the sample is heated to 725°C and annealed for 24 hours. After cooling down to room temperature, the white colour Lanarkite is achieved, which is shown in Fig.1 (d). Using these two precursors i.e., $Cu_3P$ and $Pb_2SO_5$, in 6:5 stoichiometric ratio, the palletised sample of LK-99 is vacuum sealed in quartz tube and heated for 10 hours at 925°C as shown in Fig.1 (e). After heat treatment, the polycrystalline small chunks of final product LK-99 are obtained, which are shown in Fig. 1(f). A possible chemical reaction leading to LK-99 synthesis is; $6Cu_3P + 5Pb_2SO_5 = Pb_9CuP_6O_{25} + 5Cu_2S + Pb + 7Cu$.

During the whole synthesis process, the X-ray diffraction (XRD) spectra of all finely crushed powders of precursors were performed, and the results were compared to the JCPDS data. The PXRD measurements were performed by a Rigaku-Miniflex-II table top XRD equipped with a Cu-$K_\alpha$ radiation of 1.54 Å. Fullprof software is used to perform the Rietveld refinement of obtained PXRD spectra. The sample is tested for diamagnetism by magnetic levitation of the sample. Further, for the robust test of superconductivity, magnetization measurement is done on SQUID magnetometer at 280K.

To determine the correlation effects induced due to Cu doping in Lead Oxi-Apatite, we have performed first principle calculations. For this we use first principle methods based on Density functional theory implemented in QUANTUM ESPRESSO to obtain the electronic band structure and density of states (DOS) of doped LK-99. A system of 41 atoms is simulated on a 4×4×5 grid of Monkhrost-Pack. The Perdew-Burke-Ernzerhof (PBE) type ultrasoft pseudopotential with generalized gradient approximation (GGA) is used to account for the electronic exchange and correlation. The wave functions are expanded in a plane wave basis with Gaussian smearing of the width 0.01. For the convergence of self-consistent calculation cutoff is $8.8\times10^{-9}$ Ry and charge cut-off of 400 Ry and wave function cut-off 50 Ry is used.

## Results and Discussion:

Fig. 2(a) shows the Rietveld refined data of recorded PXRD spectra of $Cu_3P$ using FullProf software. Red and black curves represent the experimentally observed and calculated intensity, respectively, whereas the blue curve shows the difference between the two. The Bragg's positions are represented by vertical green bars. The refinement is well fitted with by hexagonal phase with space group P63cm (185). The agreement of calculated intensity by Rietveld refinement with observed data confirms the absence of any impurity in the synthesized material. The obtained lattice parameters are a = b = 6.926(5) Å and c = 7.115(1) Å. The quality of fit in the Rietveld refinement is assessed using the parameter $\chi^2$, which has a value of 2.5; a lower $\chi^2$ value indicates a better fit between the experimental data and the refined model. Importantly, the Rietveld refinement analysis clearly shows that the synthesized $Cu_3P$ does not contain any extra peak of unreacted Copper. This observation is in contrast to our previous report[5], where we had observed the presence of unreacted Copper peaks in the PXRD pattern. This time, we used fresh chunk of Phosphorous and Cu was pre checked for any CuO contamination. The absence of unreacted Copper peaks suggests that the synthesis of $Cu_3P$ has been successful and complete, without any significant leftover from starting materials. This indicates the completion of the desired reaction and the formation of the target material. Fig. 2(b) shows the Rietveld refined data of recorded PXRD spectra of Lanarkite ($Pb_2SO_5$) using FullProf software. The Rietveld refinement of the PXRD data is fitted with monoclinic lattice, which belongs to the P21/m space group and the obtained lattice parameters are a = 13.754(8) Å, b = 5.7046(1) Å and c = 7.072(8) Å. Through the Rietveld refinement, it is observed that both the calculated and observed data points resembles with each other, which suggests the purity of as synthesized $Pb_2SO_5$. The goodness of fit ($\chi^2$) value is 2.01, which also indicates the phase purity of our material and absence of any other phase.

After confirming, the formation of as synthesized $Cu_3P$ and $Pb_2SO_5$, both were taken in stoichiometric ratio of 1:1 as in ref. [1, 2] and vacuum sealed in quartz ampoule. After following heat treatment protocol (Fig. 1e), the polycrystalline sample of LK-99 is obtained. This sample is named as LK-99. The PXRD of as obtained polycrystalline LK-99 samples is shown in Fig. 2(c). A few small peaks are observed as well, which may belong to the by-product of synthesized LK99. It is worth noting, that the level of foreign phases, in particular the $Cu_2S$ is relatively much smaller than the reported ones[1,2,4,5,18]. The goodness of Reitveld fitting i.e. ($\chi^2$) value is 3.46, which is reasonable. In the bottom of Fig. 2(c), the simulated XRD pattern of Apatite structure is shown, and the inset shows the top view of the Apatite structure. All major peaks of LK-99 match with the simulated XRD of Apatite structure. The Rietveld refined lattice parameters of LK-99 are a = b = 9.851(4) Å and c = 7.437(9) Å. There is a slight shrinkage as compared to that of Lead Apatite structure, which indicated the substitution of Copper atoms to the atomic sites of Lead atoms[1]. The lattice parameters for pure Lead Apatite are known[19] to be a = b = 9.865(3) Å and c = 7.431(3) Å. There is a slight reduction of 0.18% in volume, which is in accordance with contraction of unit cell parameters due to Pb site Cu substitution. The small intensity of bi-product peaks of Cu (#), $Cu_2S$ (*) and Pb (+) are marked in PXRD of presently studied LK-99 in Fig. 2(c). All the refined lattice parameters for used precursors ($Cu_3P$, $Pb_2SO_5$) and resultant LK-99 are shown in Table 1. From the XRD refinement of present LK-99, it can be concluded that the same is almost pure phase. However, further characterization is required to investigate the substitution of Cu atoms on Pb(2) atomic sites, say by synchrotron radiation. Within XRD limits, one can safely conclude that the studied LK-99 is near phase pure, with slight reduction in its volume in comparison to Lead Apatite, indicating possible Cu substitution at Pb site. Till now, the story line seems to be as good as being reported[1,2] for superconducting LK-99.

Next, we examine the acquired LK-99's superconducting characteristics. As our first test out of curiosity, we checked to see if a permanent magnet is levitated over the obtained LK-99 sample or the vice versa. Fig. 3 depicts a tiny sample piece sitting perfectly still over a permanent magnet. This demonstrates that the obtained LK-99 sample does not show any signature of levitation. Further, the magnetic properties are studied by performing the isothermal magnetization (MH) measurements at 280K, by using SQUID magnetometer. Fig. 4(a) shows the linear dependence of magnetization on the applied filed with negative slope. Fig. 4(b) shows the magnetization variation with respect to temperature (M-T profile) for our synthesised LK-99 with Field Cool (FC) and Zero Field Cool (ZFC) protocol. The M-H and M-T profile for LK-99 is similar to that of $NbAs_2$ which is a diamagnetic material[20]. Thus, magnetization measurement shows that the current sample exhibits diamagnetic behaviour and this confirms the absence of signature of superconductivity in the as synthesized LK-99 sample. In conclusion, Fig. 3 and 4 show that we have not yet been able to substantiate the assertions made in ref. [1,2] about the superconductivity of LK-99 at ambient temperature. As far as magnetic properties of LK-99 are concerned, superconductivity above 400K[1,2], 280K paramagnetic[5], ferroic nature[18] and simple diamagnetism in our present sample are reported here. In terms of phase purity, the presently studied LK-99 seems to be better than the earlier reported ones[1,2,5,6,18]. Although, the contraction in cell volume of presently studied LK-99 indicates towards the substitution of Cu at Pb site in Lead Apatite structure, yet superconductivity is not observed.

Fig. 5(a) shows the calculated DOS plot of the Lead Apatite and LK-99. A finite value of DOS shows that the doping of Cu introduces the bulk conduction channel in LK-99. Although, the parent compound Lead Apatite is insulating. Fig.5 (b) show the optimised path chosen for electronic band structure calculation in the hexagonal first Brillouin zone. Fig. 5(c) shows the calculated band structure in the first Brillouin zone of Lead Apatite and LK-99 along the high symmetric path Γ−M−K− Γ−A−L−H−A. The bulk electronic band structure of the Lead Apatite clearly shows that it is insulator. However, the electronic band structure of the LK-99 shows that there are bands close to Fermi Level crossing the same. This indicates the possible transition of LK-99 from an insulator towards conductor. It is clear that correlation effects are enhanced due to doping of Cu. Finally, although, our computational calculations support the mechanism presented in ref. [1,2,9], but the experimental results do not support the claim of bulk superconductivity in LK-99.

## Conclusion:

Summarily, we have synthesized phase pure precursors of LK-99, as confirmed by the Rietveld analysis of PXRD data. Further, we report successful synthesis of LK-99, following the same method as suggested in ref. [1,2]. The PXRD data and its Rietveld refinement of LK-99 show the shrinkage in lattice parameters, leading to a 0.18% volume contraction as compared to the parent compound Lead Apatite. But the intriguing superconductivity as claimed in ref. [1,2] appears to be elusive. The as prepared sample does not show any levitation on a permanent magnet. The isothermal magnetization measurements at 280K and the M-T measurement, show that the prepared sample, is though diamagnetic, but without any signatures of superconductivity. Although, computational results support possible superconducting mechanism in LK-99, the experimental results reveal no superconductivity in LK-99.

## Acknowledgement:

Authors would like to acknowledge the keen interest of Prof. Achanta Venu Gopal, Director CSIR-NPL in superconducting materials research. Dr. Pallavi Kushvaha is acknowledged for providing the MPMS based magnetization measurements for our sample The motivation and encouragement of Prof. G. Baskaran (IMSc/IITM) and Prof. D.D. Sarma (IISc) has been very instrumental in carrying out this research. VPS Awana acknowledges various fruitful discussion with Prof. I. Felner from Hebrew Univ. Jerusalem, concerning the observation of possible superconductivity in LK-99. The research is supported by in house project OLP-230232.

**Conflict of Interest statement:** Authors have no conflict of interest.

## Table-1

Parameters obtained from Rietveld refinement:

| | a (Å) | b (Å) | c (Å) | α | β | γ | $\chi^2$ |
|---|---|---|---|---|---|---|---|
| **$Cu_3P$** | 6.926(5) | 6.926(5) | 7.115(1) | 90 | 90 | 120 | 2.5 |
| **$Pb_2SO_5$** | 13.754(8) | 5.704(1) | 7.072(8) | 90 | 115 | 90 | 2.01 |
| **LK-99** | 9.851(4) | 9.851(4) | 7.437(9) | 90 | 90 | 120 | 3.46 |

# Figure Captions:

**Figure 1**: **(a)** Heat treatment diagram for $Cu_3P$ and **(b)** obtained product after heat treatment in sealed quartz tube. **(c)** Heat treatment diagram for $Pb_2SO_5$ and **(d)** obtained product after heat treatment in open alumina crucible. **(e)** Heat treatment diagram for LK-99 synthesis and **(f)** small chunks of polycrystalline sample.

**Figure 2**: **(a)** Rietveld refined PXRD pattern of synthesized polycrystalline $Cu_3P$. **(b)** Rietveld refined PXRD of synthesized polycrystalline $Pb_2SO_5$. **(c)** Shows the Rietveld refined PXRD of synthesized polycrystalline LK-99 along with Reitveld refinement, and the bottom shows the simulated XRD pattern for Apatite structure, the inset shows the top view of Apatite crystal structure.

**Figure 3:** The picture of the LK-99 sample siting ideal over the permanent magnet at room temperature.

**Figure 4:** **(a)** The isothermal magnetization (MH) of as synthesized LK-99 samples at 280K. **(b)** Shows the M-T curve for 500 Oe revealing diamagnetism in LK-99.

**Figure 5:** **(a)** Shows the DOS of plots for LK-99, **(b)** Shows the k-path along which electronic band are calculated, **(c)**Shows the electronic band structure of Lead Apatite **(d)** Shows calculated bulk electronic band structure of LK-99, where bands are crossing the fermi level.

## References:

1. Lee, S., Kim, J. H., & Kwon, Y. W., The Firs Room-Temperature Ambient-Pressure Superconductor *arXiv:2307.12008 (2023).*
2. Lee, S., Kim, J., Kim, H. T., Im, S., An, S., & Auh, K. H., Superconductor $Pb_{10-x}Cu_x(PO_4)_6O$ showing levitation at room temperature and atmospheric pressure and mechanism *arXiv:2307.12037 (2023).*
3. Lee, S. *et. al.,* Consideration for the development of room-temperature ambient-pressure superconductor (LK-99). *J. Korean Cryst. Growth Cryst. Technol.*, **33**, 61 (2023).
4. Liu, L., Meng, Z., Wang, X., Chen, H., Duan, Z., Zhou, X., Yan, H., Qin, P. and Liu, Z., Semiconducting transport in $Pb_9Cu(PO_4)_6O$ sintered from $Pb_2SO_5$ and $Cu_3P$. *arXiv:2307.16802 (2023).*
5. Kumar, K., Karn, N. K., & Awana, V. P. S., Synthesis of possible room temperature superconductor LK-99: $Pb_9Cu(PO_4)_6O$. *arXiv:2307.16402(2023).*
6. Hou, Q., Wei, W., Zhou, X., Sun, Y., & Shi, Z., Observation of zero resistance above 100 K in $Pb_{10-x}Cu_x(PO_4)_6O$. *arXiv:2308.01192 (2023).*
7. Wu, H., Yang, L., Xiao, B., Chang, H., Successful growth and room temperature ambient-pressure magnetic levitation of LK-99. *arXiv:2308.01516 (2023).*
8. Kim, H. T., Room-temperature-superconducting Tc driven by electron correlation. *Sci. Rep.* **11**, 10329 (2021).
9. Baskaran, G., Broad Band Mott Localization is all you need for Hot Superconductivity: Atom Mott Insulator Theory for Cu-Pb Apatite. *arXiv:2308.01307 (2023).*
10. Baskaran, G., Impurity band Mott insulators: a new route to high Tc superconductivity., *Sci. Technol. Adv. Mater.* **9**, 044104 (2009).
11. Cabezas-Escares, J., Barrera, N. F., Cardenas, C., & Munoz, F., Theoretical insight on the LK-99 material. *arXiv:2308.01135 (2023).*
12. Kurleto, R., Lany, S., Pashov, D., Acharya, S., van Schilfgaarde, M., & Dessau, D. S., Pb-Apatite framework as a generator of novel flat-band CuO based physics, including possible room temperature superconductivity. *arXiv:2308.00698 (2023).*
13. Si, L., & Held, K., Electronic structure of the putative room-temperature superconductor $Pb_9Cu(PO_4)_6O$. *arXiv:2308.00676 (2023).*
14. Lai, J., Li, J., Liu, P., Sun, Y., & Chen, X. Q., First-principles study on the electronic structure of $Pb_{10-x}Cu_x(PO_4)_6O$ (x=0,1). *arXiv:2307.16040 (2023).*
15. Oh, H., & Zhang, Y.-H., S-wave pairing in a two-orbital t-J model on triangular lattice: possible application to $Pb_{10-x}Cu_x(PO_4)_6O$. *arXiv:2308.02469 (2023).*
16. Tao, K., Chen, R., Yang, L., Gao, J., Xue, D., & Jia, C., The Cu induced ultraflat band in the room-temperature superconductor $Pb_{1-x}Cu_x(PO_4)_6O$ (x = 0,0.5). *arXiv:2308.03218 (2023).*
17. Hlinka J., Possible Ferroic properties of Copper-substituted lead phosphate Apatite. *arXiv:2308.03691(2023).*
18. Guo, K., Li, Y., & Jia, S., Ferromagnetic half levitation of LK-99-Like synthetic samples. *arXiv:2308.03110 (2023)*
19. Krivovichev, S.V., Burns, P.C., Crystal chemistry of Lead Oxide Phosphates: Crystal Structures of $Pb_4O(PO_4)_2$, $Pb_4O_8(PO_4)_2$ and $Pb_{10}(PO_4)_6O$. *Z. Kristallogr.* **218**, 357 (2003)
20. Peramaiyan, G., Sankar, R., Muthuselvam, I.P. & Lee, W. L., Anisotropic magnetotransport and extremely large magnetoresistance in $NbAs_2$ single crystals. *Sci. Rep.* **8**, 6414 (2018).

Fig.1:

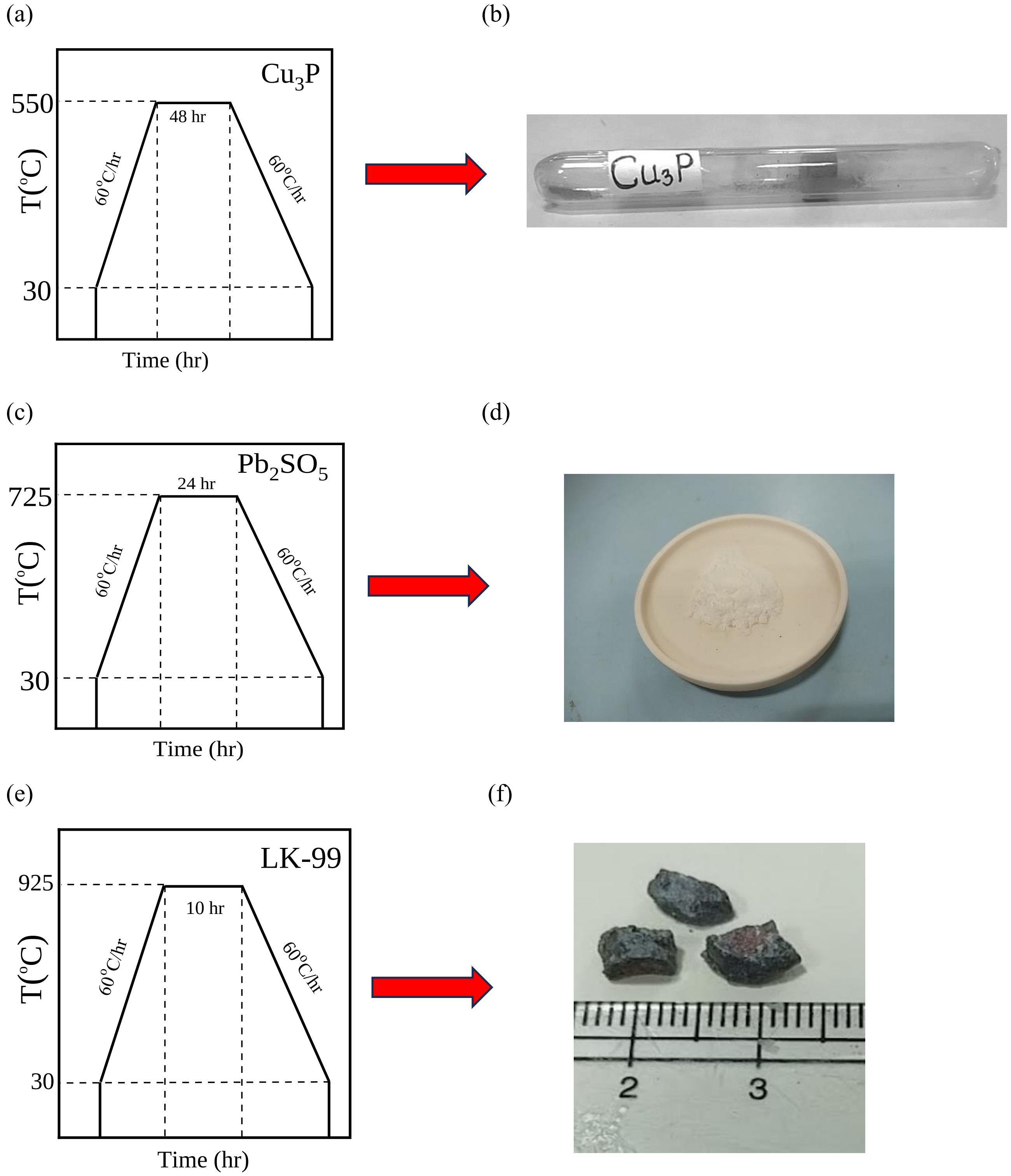

Fig.2:

(a)

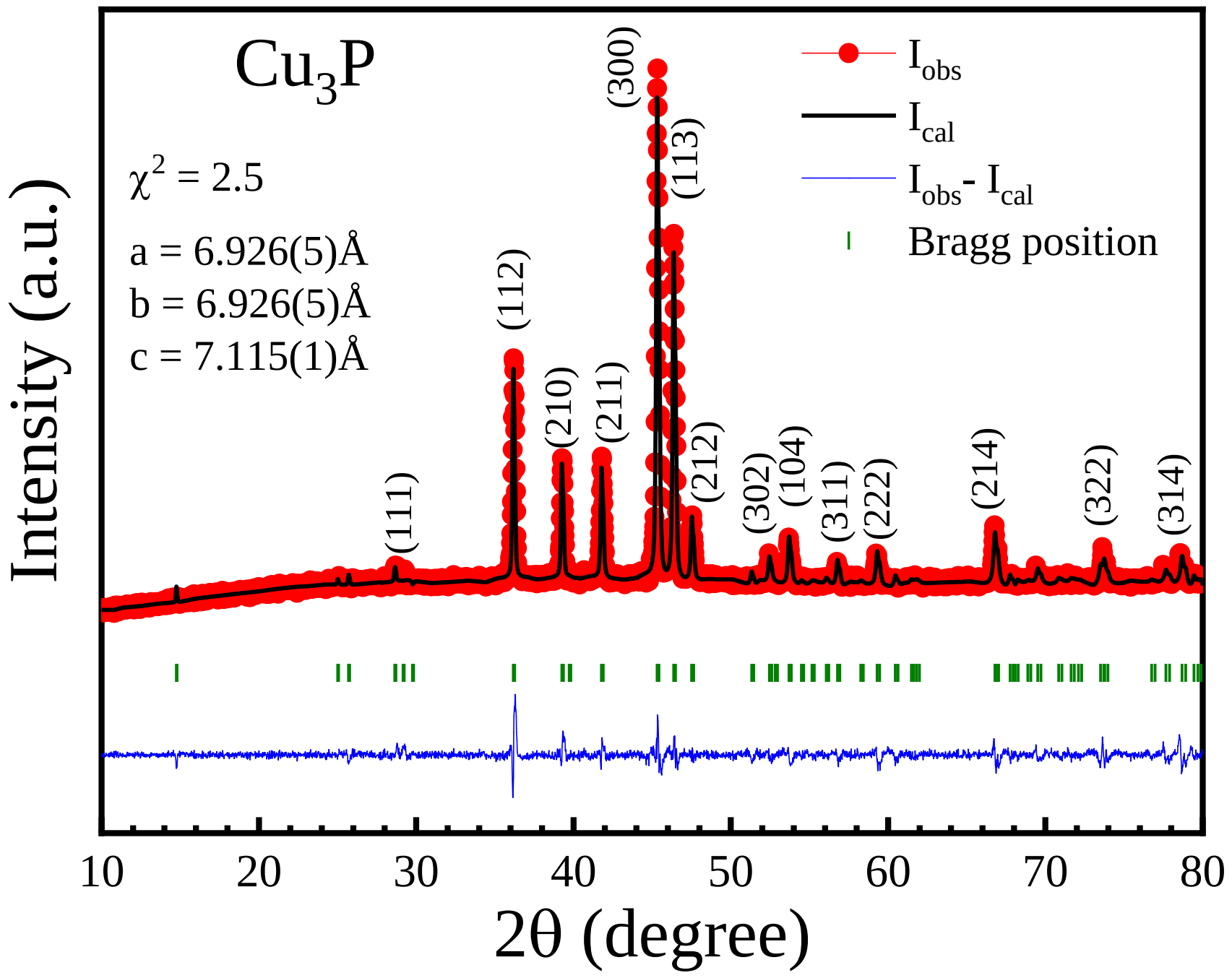

(b)

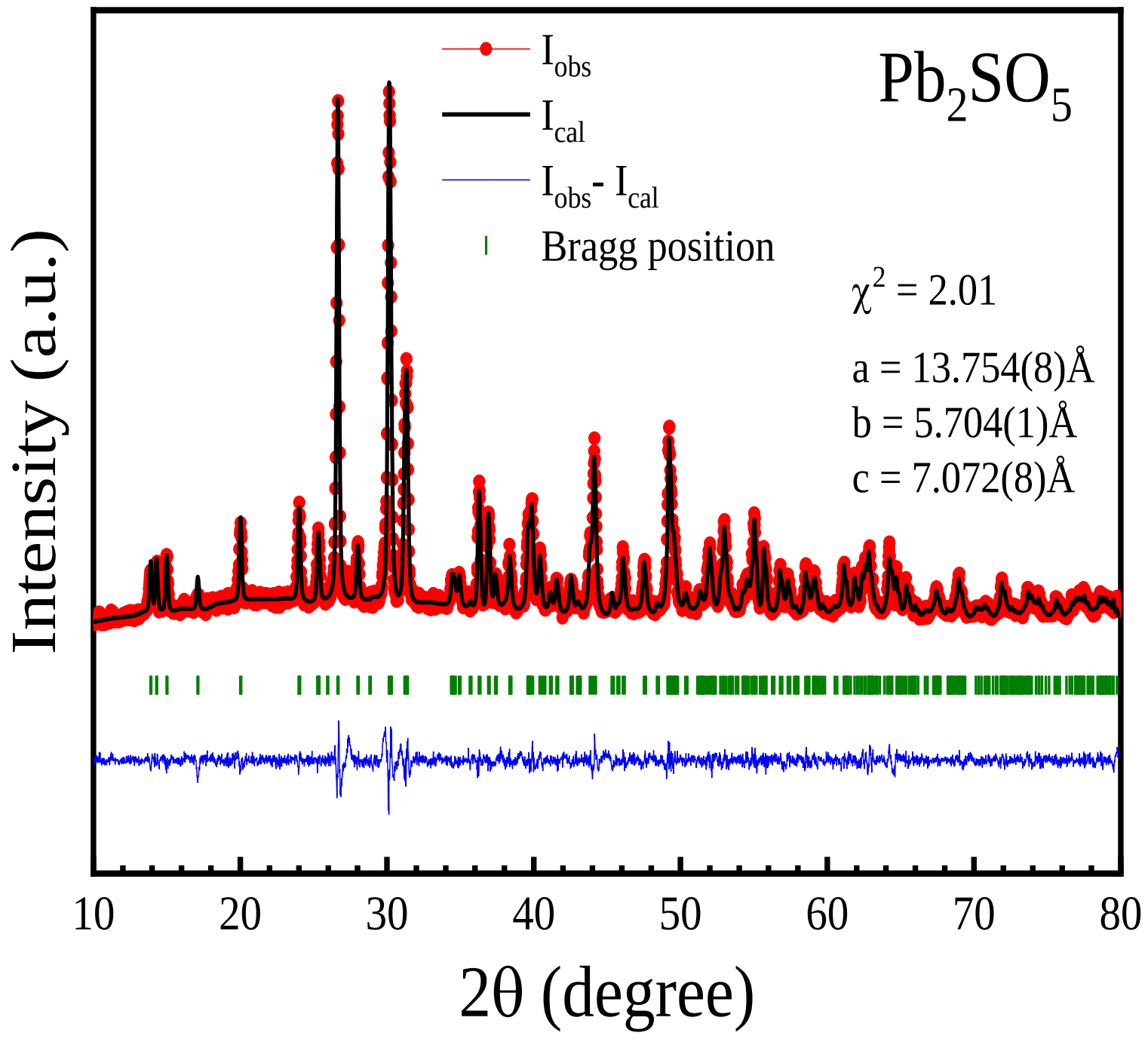

(c)

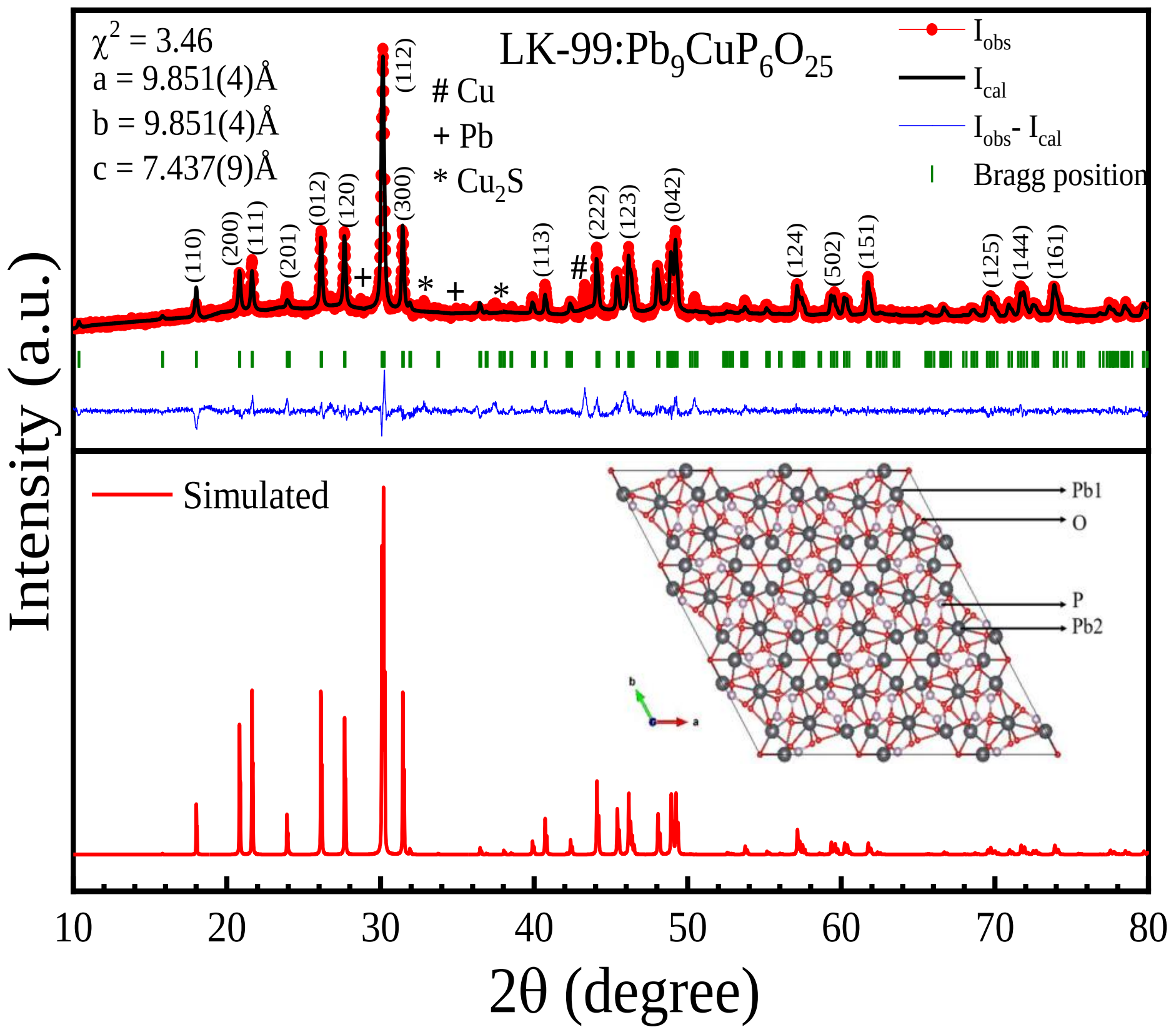

Fig.3:

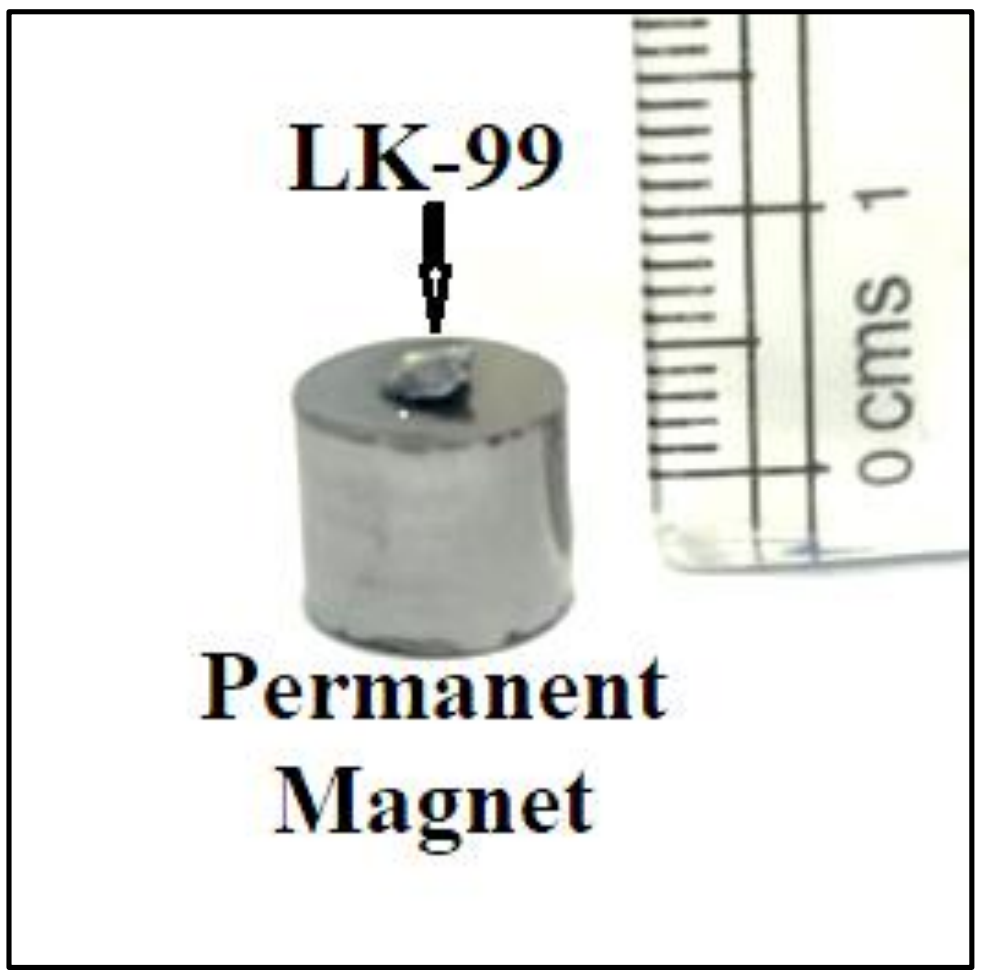
LK-99
Permanent
Magnet
0 cms 1

Fig.4: (a)

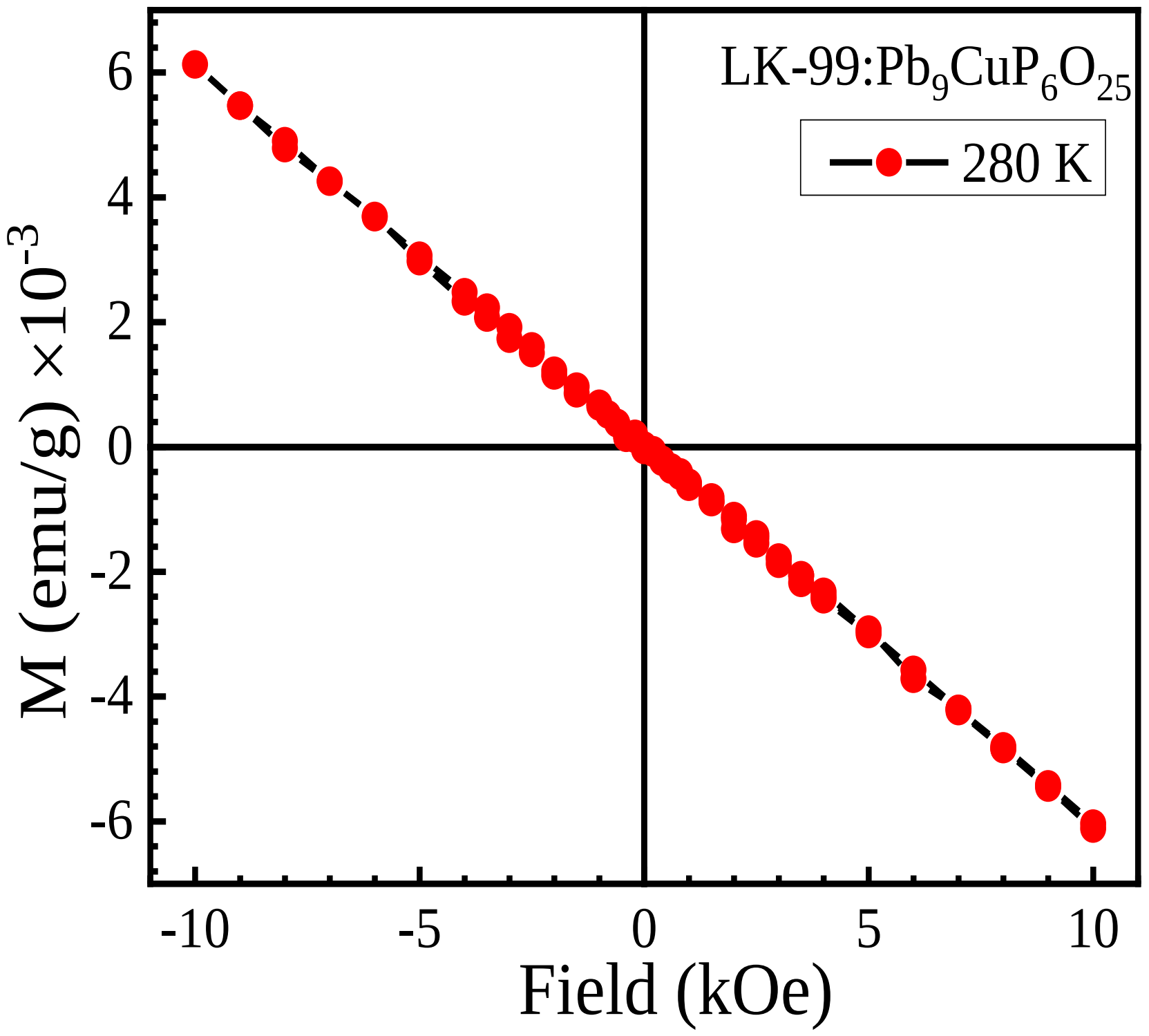

Fig. 4(b)

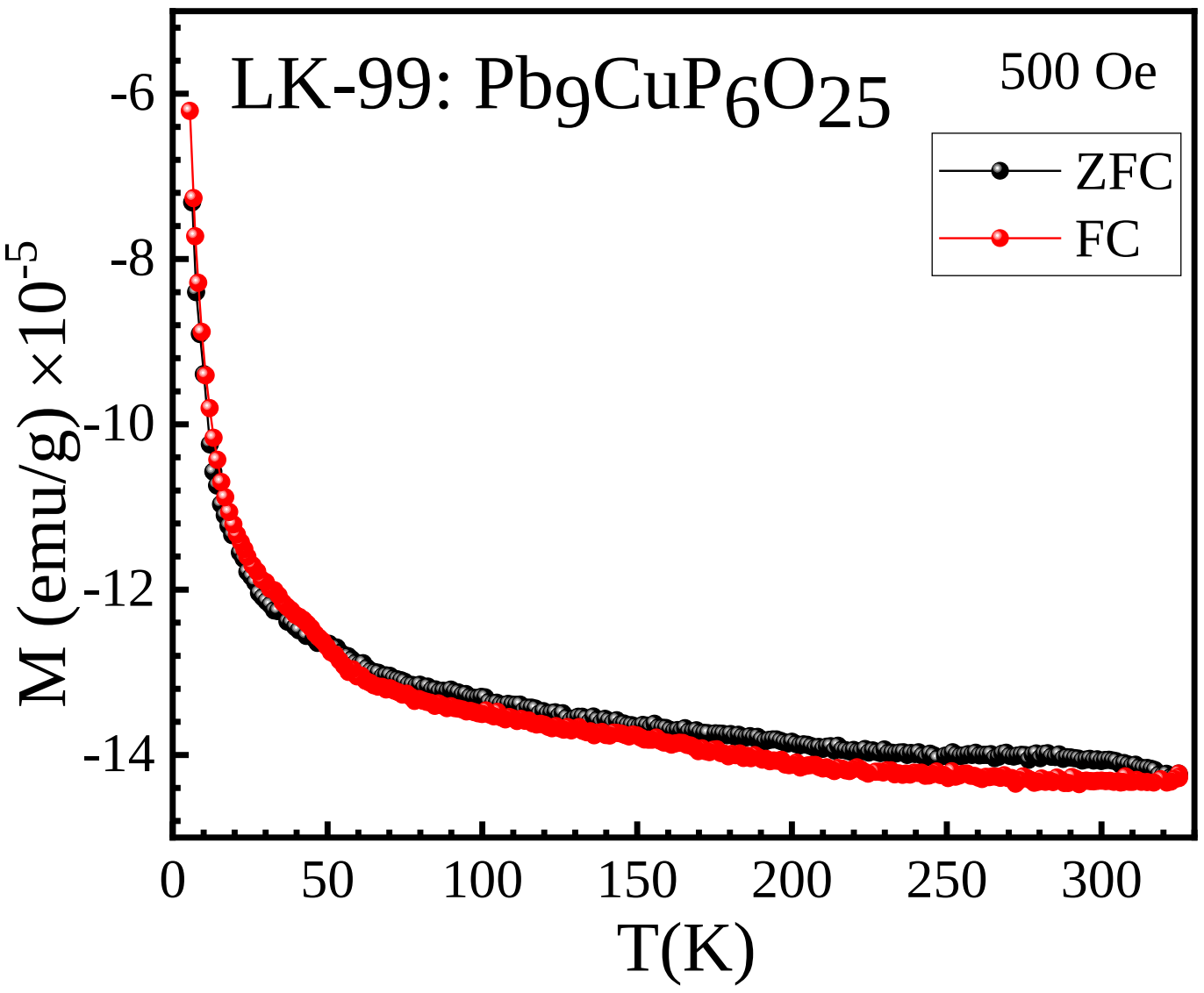

Fig.5: (a)

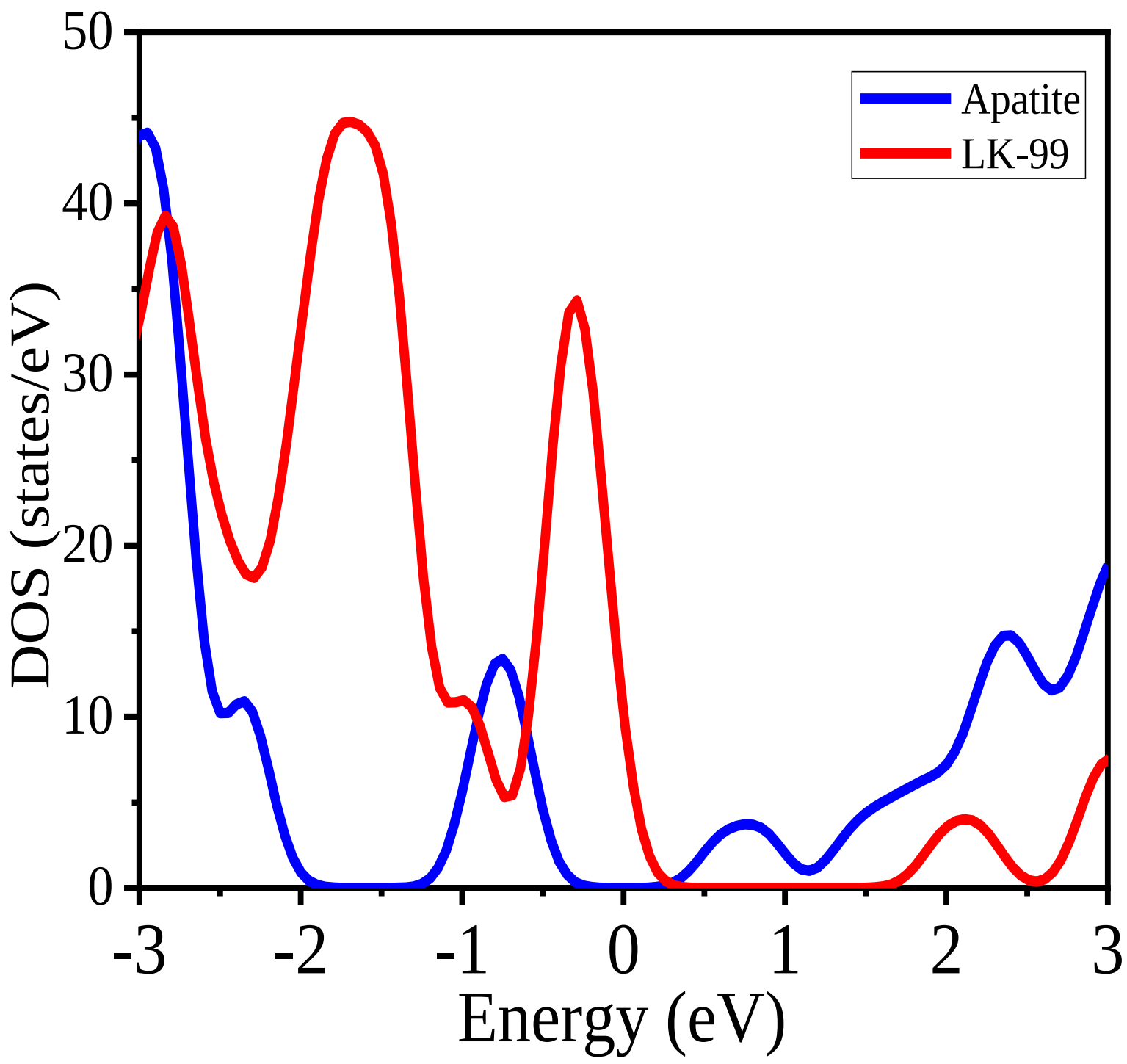

Fig.5: (b)

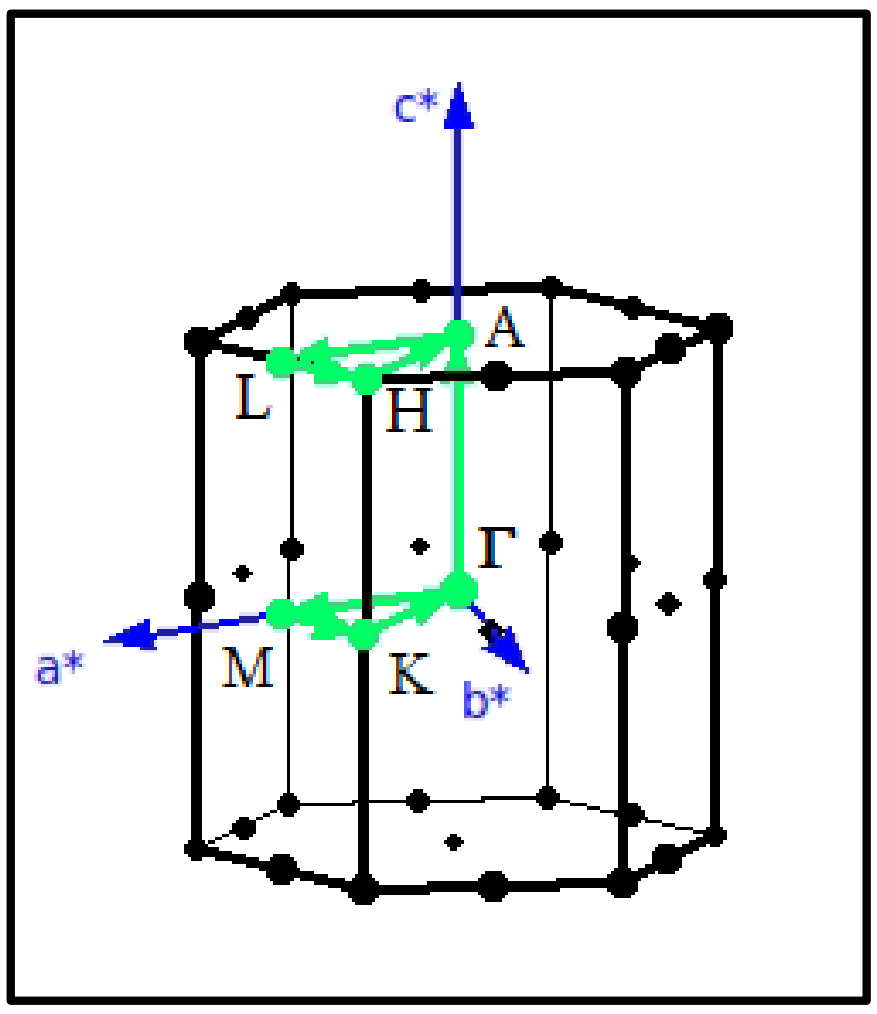

Fig.5: (c)

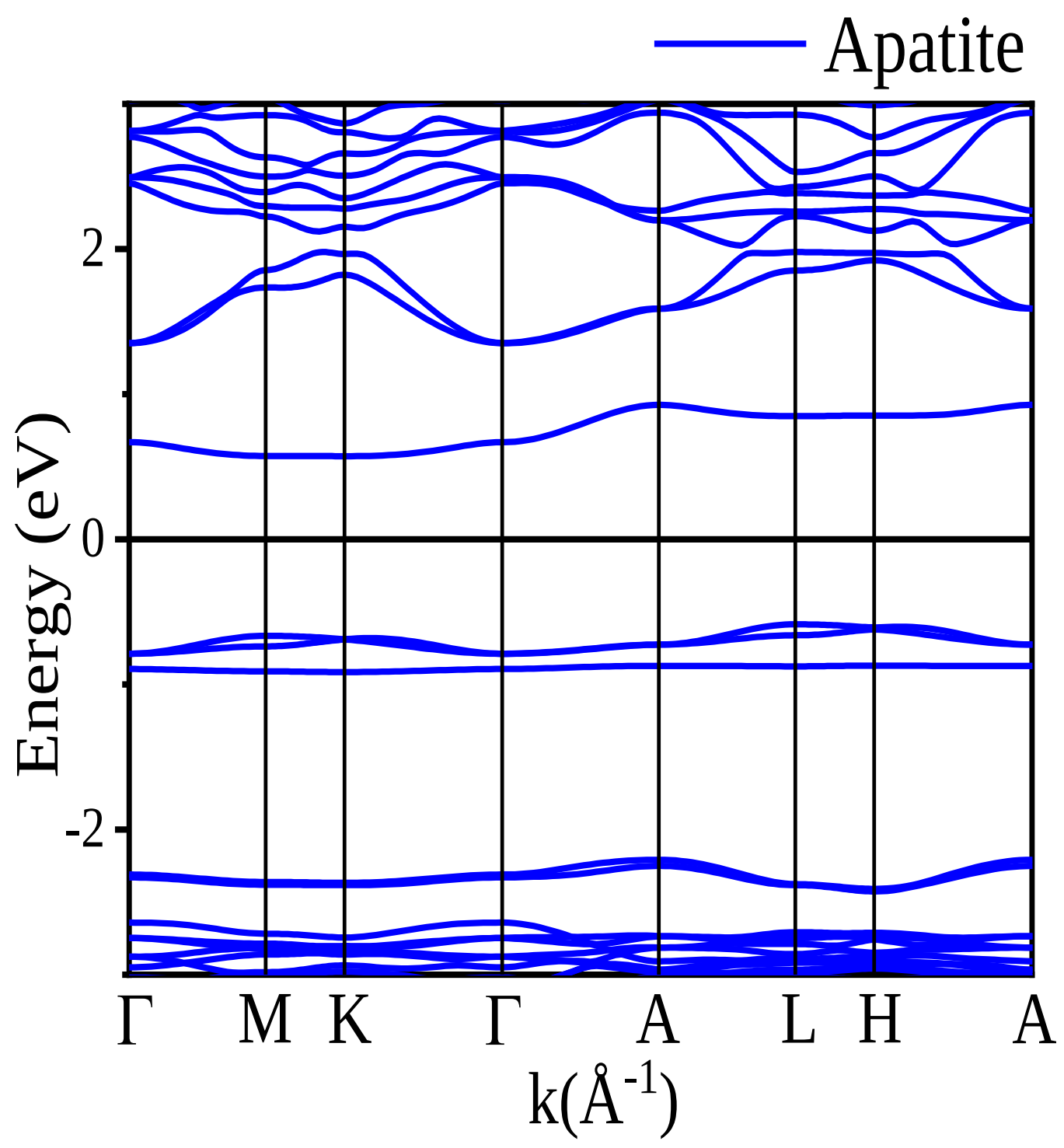

Fig.5: (d)

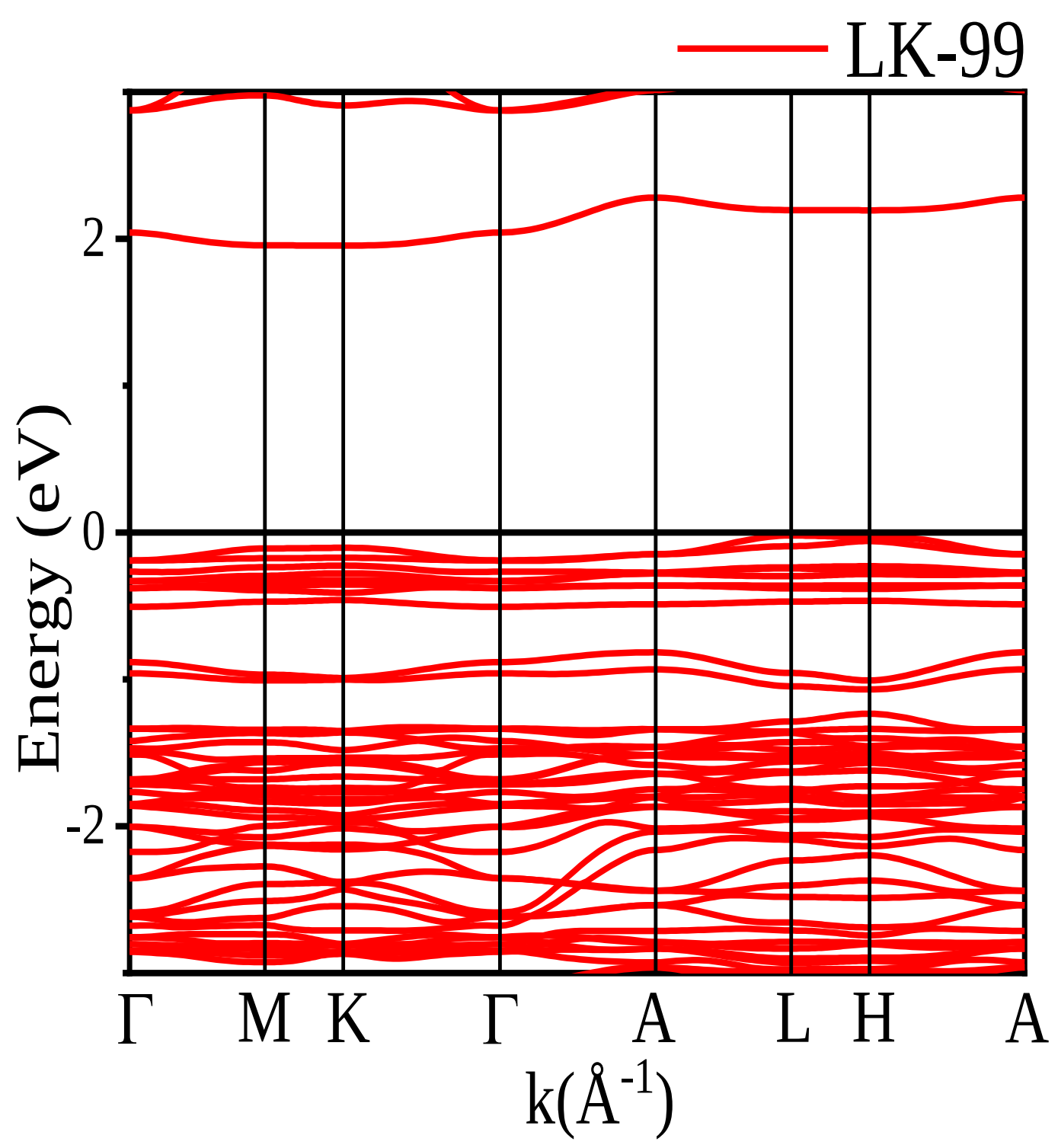